\documentclass[showpacs,preprintnumbers,amsmath,amssymb]{revtex4}
\usepackage{graphicx}
\usepackage{bm}
\usepackage{setspace}

\begin{document}
                                                                       
\preprint{APS/123-QED}
                                                                                
\title{Stokes experiment in a liquid foam.}
                                                                                
\author{I. Cantat}
\email{isabelle.cantat@univ-rennes1.fr}
\affiliation{GMCM, UMR CNRS 6626, 35042 Rennes cedex, France}
                                                                                
\author{O. Pitois}
\affiliation{LPMDI, UMR CNRS 8108, 77454 Marne-la-Vall\'ee Cedex 2, France}

\date{\today}
                                                                                
\begin{abstract}
The paper reports on the quasi-static steady flow of a dry liquid foam around a fixed spherical bead, few times larger than the typical bubble size. The force exerted on the bead is recorded with a precision and a time resolution  large enough to show the succession of elastic loading of the foam, separated by sudden force drops. The foam structure 
is observed by direct light transmission, synchronized with the force measurement, thus allowing to correlate 
the  plastic events with the force variations. Scaling laws for the force signal as a function of the bubble size are detailed and interpreted with a simple elasto-plastic model. The spatial distribution of the plasticity is strongly localized in the first bubbles layers around the bead and the average size of bubble rearrangements increases with the corresponding force jump amplitude.

\end{abstract}
                                                                                
\pacs{83.80.Iz,83.60.La,62.20.Fe}

\title{Stokes experiment in a liquid foam.}

\maketitle

\newpage
\section{Introduction}

At small deformations, a dry liquid foam behaves like an incompressible soft elastic solid. The bubbles elongate and the amount of gas/fluid interface increases : the foam stores elastic energy and develops a restoring force. At larger  strain,  the deformed structure becomes instable, the bubbles slide over each other and thus recover a more isotropic shape
(for a review see \cite{kraynik88,weaire,hohler05}).
Among the complex fluids, the foam is probably the material for which the elementary processes leading to 
elasto-plasticity are the best understood, and the numerical simulations the more accurate.  The involved scale is the millimetric bubble scale, at which the structure and/or the velocity field can be monitored with various
resolutions and various techniques, among which X-ray tomography\cite{lambert05}, diffusing wave spectroscopy (DWS) \cite{durian91}, optical tomography \cite{monnereau98b,rouyer03,vandoornum06}, magnetic resonance imaging (MRI) \cite{prause95,coussot02,rodts05}. 
Liquid foams are thus very promising model systems to understand the coupling between the local organization and 
the macroscopic rheological response. Generalization is expected to other complex fluids, as colloidal suspensions, slurries, pastes or clays, in which the structure scale is nanometric. 

Most of the 
experimental results on 3D foams concern homogeneous simple shear geometries.  The intermittency, the 
stress and strain inhomogeneities and their tensorial nature are crucial parameters that cannot be 
tackle with classical rheometers. 
In contrast, the Stokes experiment, {\it ie} the motion of a rigid obstacle in a liquid system, induces an inhomogeneous stress in the material and thus provides very complementary information. Some previous works report on the relation between the velocity of a foam flow and the force it exerts on an obstacle in 2D \cite{dollet05, dollet05b} or in 3D \cite{cox00,debruyn04,cantat05b}. Nevertheless, the force fluctuations, correlated to the direct observation of the rapid bubbles rearrangements responsible for the plastic transformations 
 are presented here for the first time. 

Our experimental set-up, detailed in section \ref{exp_meas}, imposes a steady foam flow around a bead. The force exerted on the bead is recorded with a precision scale under which the bead is rigidly fixed, as already used in \cite{cantat05b}. A camera is synchronized with the scale and records the images of the foam structure. In section \ref{shear_mod}, the foam shear modulus and its plastic threshold are deduced from the force signal and compared to predictions obtained with a continuous elasto-plastic medium. The force fluctuations are detailed in section \ref{force_jump} and the correlation with 
the magnitude of the plastic events, obtained from the image processing, is demonstrated. Finally, the spatial distribution 
of the foam structure rearrangements is discussed in section \ref{spatial_dis}.

\section{Experimental measurement}
\label{exp_meas}

A monodisperse dry foam, with millimetric bubbles, is produced {\it in situ} in a transparent 
$10 \times 10 \times 20$ cm Plexiglas box by nitrogen blowing in a solution of SDS (sodium dodecyl sulfate) 3g/L and dodecanol 0.01g/L. The bubble size can be varied by changing the gas flux and the injector needle diameter.
The surface tension of the solution is $\gamma = 37$mN/m, its viscosity  $ \eta = 10^{-3}$ kg m$^{-1}$ s$^{-1}$ and its density $\rho = 10^3$ kg/m$^3$. Once the cell is filled, the box is closed to avoid evaporation and films breakage,
and the remaining solution is removed from the cell bottom. 
As discussed in \cite{cantat05b}, aging processes (drainage and coarsening) strongly perturb the measurements so we let the foam draining
 during half an hour, until the liquid 
fraction $\phi$ becomes constant and of the order of 1 or 2 $\%$.
The coarsening is slowed down by use of $C_6 F_{14}$ traces added to the nitrogen. The experiments presented in this paper have  been done between half an hour and two hours after the foam formation in section \ref{shear_mod} and between half an hour and one hour in section \ref{force_jump}.

All the bubbles have the same volume, with a dispersion smaller than $10 \%$. Its value, which is our main control parameter,
can be obtained by taking a picture of a spherical bubble rising in the liquid phase, during the foaming process. Nevertheless, it is more convenient to measure the average distance $d$ between two horizontal films within the bubble column at the box corner. The first method has thus been used only once to determine the relation between $d$ and  the 
bubble volume, or, as used classically in the literature, the radius $r_V$ of the sphere of same volume.
We found 
\begin{equation}
r_V= 0.54 \, d \pm 5\% \, , 
\label{rVded}
\end{equation} 
which will allow quantitative comparisons with previous works.

\begin{figure}[h]
\includegraphics[width=10cm]{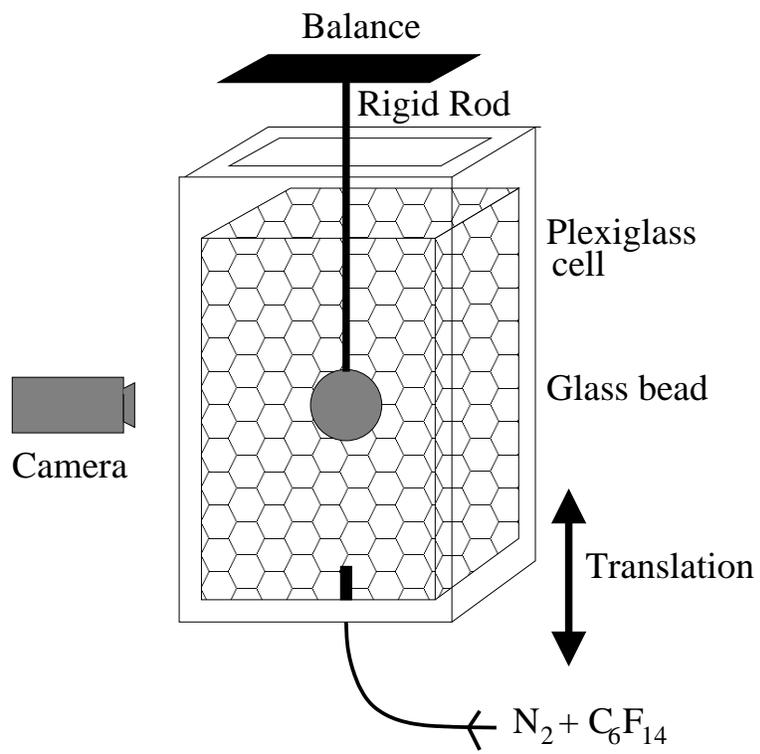}
\caption{Cantat, Physics of Fluids.}
\label{set-up}
\end{figure}

A glass bead of radius $R=0.5$cm is maintained at a fixed position, in the middle of the horizontal box cross-section, 
 by a rigid glass rod. The sphere and the rod are entirely wetted by the solution.  The box can translate vertically
at a very low velocity $V$ and thus imposes the foam motion around the immobile 
bead (see Fig.\ref{set-up}). 
The rod is fixed below a precision balance (Sartorius BP 211D) and 
the force exerted by the foam on the bead is recorded every 0.25s with a precision of 1 $\mu$N. The reference force is taken without foam in the box.

\begin{figure}[h]
\includegraphics[angle=-90,width=11cm]{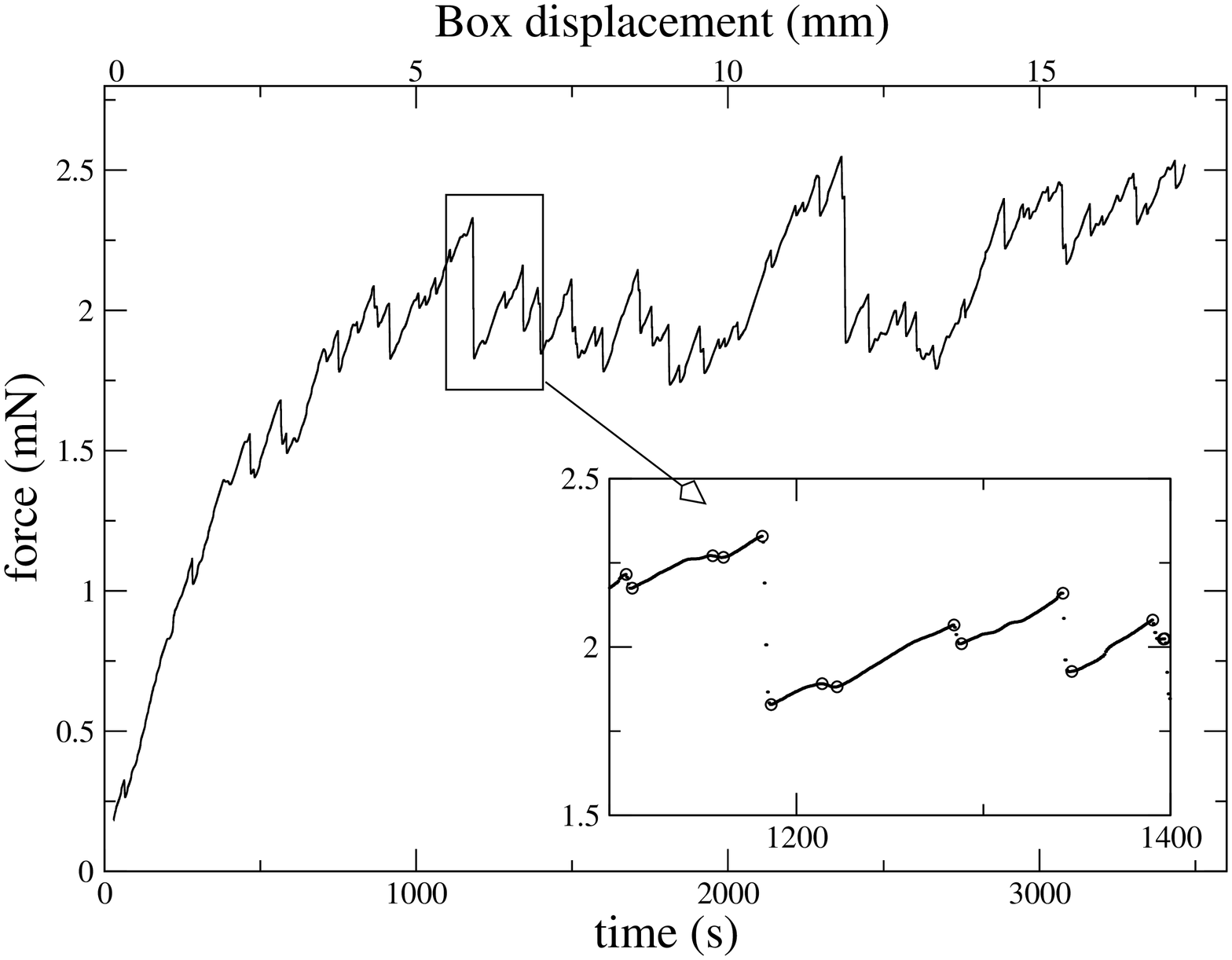}
\caption{Cantat, Physics of Fluids.}
\label{allure_force}
\end{figure}

When the box begins to move, the force firstly increases during a transient, which is disregarded in this study (see Fig.\ref{allure_force}). The end of this transient is
estimated roughly as the time at which the force reaches its mean value and the data collected before this time are suppressed from the statistics. Then we observe a succession 
of sharp force decreases (plastic events) separated by smooth force increases (elastic loading) during which the force varies linearly with the box displacement x, even for the longest force increases observed (see Fig.\ref{allure_force}).
The beginning and the end of the elastic loading indexed by $i$ are denoted respectively by $t_{min,i}$ and $t_{max,i}$ and are extracted from the 
raw force data with the following method. After the jump $i-1$, a new loading $i$ begins at the first time verifying $F(t+dt) > F(t) + \epsilon$, with dt=0.25s. $F(t_{min,i})$ is denoted by $F_{min,i}$. The parameter $\epsilon$ is fixed slightly above the noise level at  $\epsilon= 5 \mu$N.  The end of the loading $i$ (or equivalently the beginning of the jump $i$) occurs at the first time $t_{max,i}$
at which $F(t+dt) < F(t) - \epsilon$, and $F_{max,i}=F(t_{max,i})$. If the foam moves upwards, we replace F by -F in the data processing. From this list of force maximum and minimum, we compute the jump amplitude 
$\delta F_i=F_{max,i-1}-F_{min,i}$, the rate of force
increase during elastic loading $dF/dt)_i=(F_{max,i}-F_{min,i})/(t_{max,i}-t_{min,i})$, the waiting time between two jumps $t_{w,i}= t_{max,i}-t_{max,i-1}$ and the jump duration $t_{d,i}= t_{min,i}-t_{max,i-1}$.
The time resolution is much smaller than the typical force jump duration (of the order of 1.5 s as demonstrated in section \ref{force_jump}) and is not a limiting factor. The parameter $\epsilon$ has been chosen in order to disregard the small amplitude force jumps for which the correlation with the plastic events observed
on the images disappears.  We checked carefully that it does not influence the amplitude distribution of force jumps larger than $\epsilon$. The end of the transient is not precisely determined, but the uncertainty thus induced on the mean force value is smaller than the error bar and it does not modify the other quantities.

As demonstrated on Figs. \ref{pente} and \ref{force_moyenne},
the force signal is symmetric when the displacement direction is modified and the presence of the rod seems to be without major influence. The Archimedes force, that could also break 
 the up/down symmetry, 
  is of the order of 
$4/3 \,  \pi R^3 g \rho  \phi \sim 50 \mu$N, which is much lower than the typical force measured, of the order of few mN.  
The flow occurs in a quasi-static regime, and, as checked experimentally over the range (1-50$\mu$m/s), the velocity only rescales the time unit. This is consistent with the small viscous force value estimated as follow. 
The viscous forces $f_v$ between the foam and the bead are localized along the Plateau borders and are of the order 
of $40 \gamma Ca^{2/3} \sim 20 \mu$N/m per unit length of Plateau border touching the sphere, with $Ca= \eta V / \gamma \sim 10^{-6}$ the capillary number\cite{cantat04}. The resulting force is then approximately $8 \pi R^2 f_v/ d \sim 3 \mu$N, which is negligible. Instead of $dF/dt$, the pertinent force variation is thus $dF/dx$, with $x$ the box displacement, simply obtained using the relation $dF/dt = V dF/dx$.

\begin{figure}[h]
\includegraphics[width=10cm]{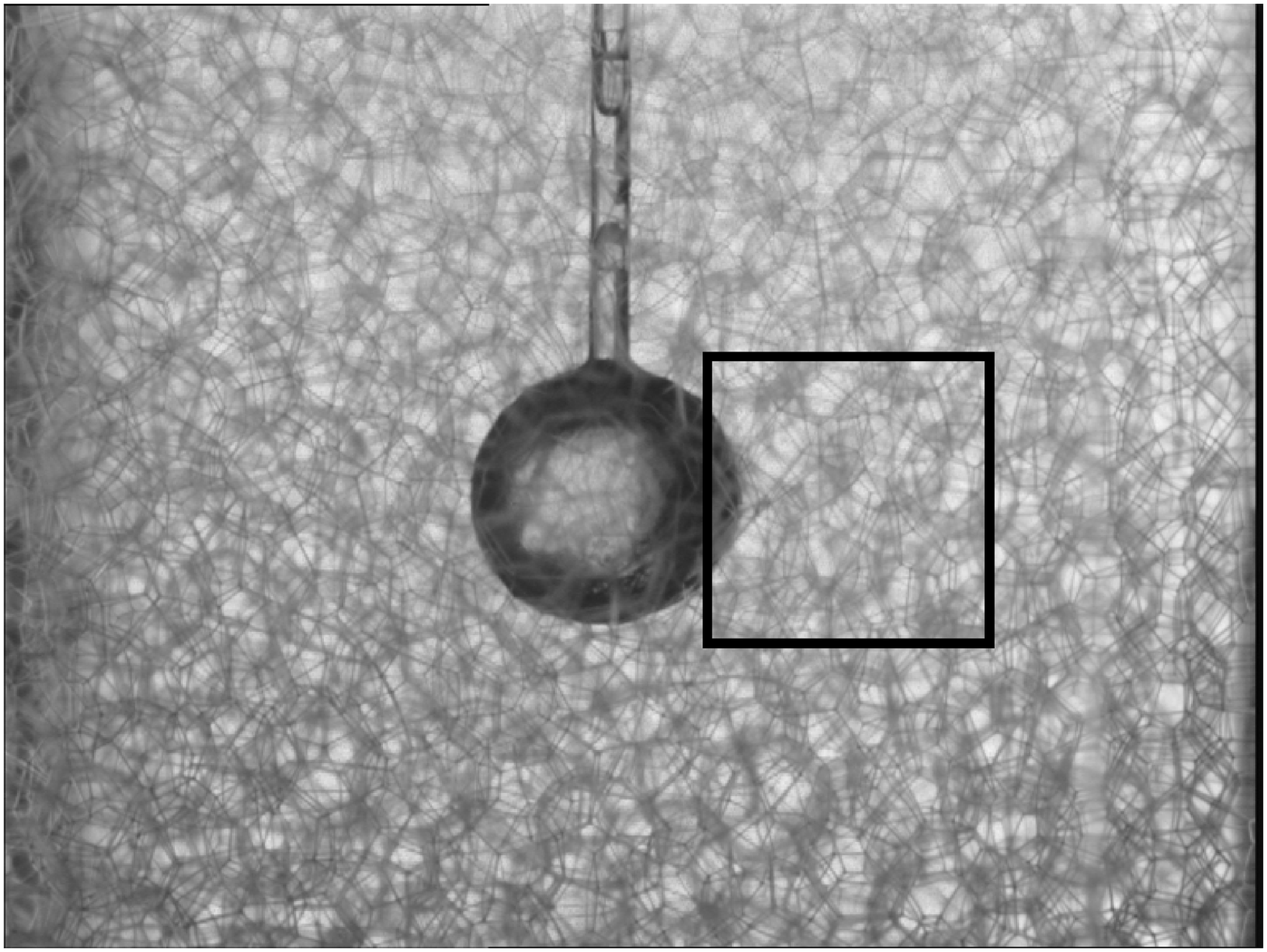}
\caption{Cantat, Physics of Fluids.}  
\label{image}
\end{figure}

The foam is lit by a cold and uniform light of the same size than the box (Schott-Fostec DDL) and the foam images are recorded 
from the opposite side with a  camera focused on the bead, at a rate of 4 images per second. The Plateau borders diffuse the light and appear in black on white, projected on a 2D image of the whole 3D foam (see Fig.\ref{image}). The slow motion of the foam at velocity $5 \mu$m/s induces 
a translation of $1.25 \mu$m between two successive images, which is much smaller than the pixel resolution ($60 \mu m $) and than the Plateau border typical radius ($a= 200-300 \, \mu$m). 
Therefore, if the foam only translates in front of the camera, 
the difference between two successive images shows very low grey levels, corresponding to the noise. In contrast, bubbles rearrangements by T1 processes, {\it ie} neighbor exchange between bubbles, induce rapid 
films motions and  lead to the typical patterns shown on Fig.\ref{diff_image}(a).
For bubbles larger than 2.5 mm, these plastic transformations can be localized automatically by the following computer image processing, made with the commercial software Visilog. 
The difference between two successive images
is binarised, keeping only the largest and smallest grey levels, smoothed on the bubble scale and binarised again. A succession of dilatation and erosion allow to keep only domains with a size comparable to one bubble size or larger.  We finally obtain domains without hole called "spots" in the following. They are approximately of the size of the few bubbles 
involved in the corresponding plastic event (see Fig.\ref{diff_image}(b)).
  
The typical duration of a plastic event is of the order the second and it is detected in average on 1 or 2 images.  The criterion to identify two successive
 spots as the same 
event is that their centers of mass are closer than the geometric average of their characteristic radius, computed from their 
areas.
 Because we have only a 2D information, it may happen
  that two distinct events lead to spots
at the same place on an image or on two successive images. Nevertheless, as discussed in section \ref{force_jump}, only 2 or 3 plastic events happen during a force jump so this spots superposition occurs rarely and the case has been neglected.
  The presence of a plastic event at a given time is well detected by our method, except for very small displacements that usually correspond to the beginning or the end of a larger event. In contrast, it is difficult to define and to quantify precisely the event size, which depends 
obviously on the grey level threshold and of the pixel resolution. 
The slightly arbitrary volume assigned to a plastic event in the data presented below is the area of the corresponding spot to the power 
$3/2$ (or, in case of an event with successive spots, the area of the largest one).  It can be used to compare plastic events between each other, but not to determine the number of elementary  T1 events involved (each T1 involving only four bubbles). 
For further experiments, the use of two synchronized crossed cameras will improve the determination of the size and position of these rapid rearrangements.  Finally, no information is obtained from the foam column in front or behind the bead.

\section{Shear modulus and plastic threshold}
\label{shear_mod}
Among other technics summarized in table \ref{table_mod}, the foam shear modulus is usually measured in a simple shear geometry, by small amplitude oscillations around an equilibrium position, as the ratio between the imposed displacement and the macroscopic stress.
 In contrast,
the Stokes geometry induces non-uniform stress and strain
in the foam, in a steady state flow. Nevertheless, assuming a purely elastic response of the medium during the smooth force increase, an expression for the shear modulus can be extracted from the comparison between the measured force and a continuum medium analytical predictions.
 
The distribution of the slopes of the force-displacement data during the elastic loading is relatively narrow (see Fig. \ref{allure_force} and \ref{pente_dis}). The mean slope was computed for each flow using a weight proportional to the duration of each slope (or, equivalently, all individual force increases are added without weight and the sum is divided by the total time 
spent in the elastic state).
The experimental data are in good agreement with the expected linear dependence with $1/d$, for a bubble size varied by more than a factor 2, and is well adjusted by the relation given on Fig.\ref{pente}. 
\begin{equation}
{dF \over dx} = {2.0 \; 10^{-3} \pm 15\%  \over d}
\label{slope_fit}
\end{equation}
with the force slope expressed in N/m and $d$ in m.
The error bar obtained from the best Gaussian fit displayed on Fig. \ref{pente_dis} and the incertitude on the fitting parameter
on Fig. \ref{pente} both lead to a precision of the order of $\pm 15\%$ 

 The force needed to displace a rigid sphere embedded in an infinite
incompressible 
continuum medium is simply, assuming a free sliding on the bead,  
\begin{equation}
F_{elas}= 4 \pi \,  G \,R\, \delta_x \; ,
\label{f_tot_elas}
\end{equation}
with R the sphere radius, $\delta_x$ the displacement amplitude and G the shear modulus of the surrounding medium
(see appendix A).  
The foam shear modulus is thus deduced from eqs.\ref{slope_fit} and \ref{f_tot_elas}.
Once scaled by $\gamma/d$ or $\gamma/r_V$ (see eq.\ref{rVded}), we obtain :
  \begin{equation} 
  G = 0.86 \,  \gamma /d \; \pm 15 \% = 0.46 \,\gamma /r_V \; \pm 20\% \, . 
 \label{young_modulus}
   \end{equation}

This value is compared to previously obtained results in the table \ref{table_mod}. 
Slip at the solid boundaries is often mentioned to explain the dispersion of the experimental results
obtained with classical rheometers. A better control is obtained with our set-up that assumes a perfectly slipping surface on the bead, which is easier to ensure than a no-slip condition. Nevertheless, the discrepancy between the various results may arise from more fundamental reasons related to geometrical effects. Our value is indeed very close to the value obtained by the indentation test \cite{coughlin96}, which is also based on a local mechanical action on the foam. 
 Numerical simulations on the Kelvin cell done by Kraynik {\it et al} demonstrate that the foam shear modulus varies between 
$0.35 \gamma/r_V$ and $0.60 \gamma/r_V$, depending on the shear orientation \cite{kraynik96}.  
Structure reorganizations around the bead, due to the many T1 induced by the flow, may lead to an important local foam anisotropy and to a decrease of the shear modulus in the direction of the applied force.

The yield stress is more difficult to define than the shear modulus. We deduced it from the mean force value, during the steady state. 
This force $F_{th}$ is found to varies linearly with $1/d$ (see Fig. \ref{force_moyenne}) and is well fitted by the relation, scaled by $R^2 \gamma/d$, 
\begin{equation}
F_{th} = 5.65 \, R^2 \gamma/d \pm 15\% \; . 
\label{seuil_exp}
\end{equation}
In the frame of the purely elastic model presented
above, the stress tensor has a simple analytical expression, detailed in Appendix A.
The shear stress, which can be more easily compared to previous values obtained in classical 
rheometers, reaches its maximal value $\sigma_{max}$ in front of the bead or at the rear : 
\begin{equation}
\sigma_{max} = 1.5 {G \delta_x \over R} = 0.75 {F_{elas} \over 2 \pi R^2} \; . 
\label{stress_elas_max}
\end{equation}
From eqs.\ref{f_tot_elas}, \ref{young_modulus} and \ref{seuil_exp}, we deduce the mean value of the effective bead displacement in the steady state (with respect to its equilibrium position, deduced from the purely elastic model) $\delta_{x,th}= 0.52 R$. Using eq.\ref{stress_elas_max}, we finally get the yield shear stress, scaled by $\gamma/d$,
\begin{equation}
\sigma_{th} = 0.65 \, \gamma/d  \pm 15\% \; = 0.35\, \gamma/r_V  \pm 20\%  \; . 
\label{yield_stress_d}
\end{equation}
Previous experimental results are very scattered for dry foams, with a measured yield stress in the range  (0.04 - 0.2 $\gamma/r_V$) (see \cite{gardiner98,rouyer05} and references therein). Gardiner {\it et al} also mention the value $\sigma_{th}= 0.56 \, \gamma/r_V$ for a very dry foam\cite{gardiner98}.  Numerical simulations by Kraynik {\it et al} on polydisperse foams
lead to $\sigma_{th} =0.18 \gamma/r_V$ with $\phi =0$.
As discussed in \cite{hohler05}, the results scattering is probably mainly due to a rapid variation of 
$\sigma_{th}$ with the liquid fraction when $\phi$ is close to 0 and to the various definitions used, which have not been demonstrated to be equivalent.
The surprisingly large value we obtain may be due to the fact that the maximal stress is only reached in a very small foam volume, which decreases the probability to found a weak structure in the 
concerned volume. Numerical simulations of a Bingham fluid or an Herschel-Bulkley fluid flowing around a sphere also leads to a relation between $\sigma_{th}$ and  $F$\cite{beris85,beaulne97,blackery97}. Anyway, as a non-slip condition is assumed on the sphere, a direct comparison is not possible.   

The maximal elongation of the first bubble layer before detaching from the sphere is another characteristic of the plastic threshold. Unfortunately the image resolution is not sufficient to obtain a direct measure. Within the continuum elastic medium approach, the maximal elongationnal stress $\sigma_{rr} = 2 G \delta_x / R$ is determined in Appendix A, using the spherical coordinates with a polar axis oriented in the direction of the bead motion. The maximal elongationnal strain $\epsilon_{rr}$ at the rear of the bead, is thus : 
\begin{equation}
\epsilon_{rr} = \sigma_{rr}/(2 G) = 0.52  \pm 10\% \; ,
\label{elong_max}
\end{equation}
The limit of stability of a 2D foam between two plates when the gap width is increased has been recently determined analytically by S. Cox {\it et al} \cite{cox02}. A n-sided bubble detaches from one 
of the two plates when the gap $g$ between the plates reaches $g(n)=1.1 \, r_V \sqrt{6/(6-n)}$, leading to the maximal elongation
for $n=5$ of the order of $(g(5)-2 r_V)/(2 r_V) \sim  0.35$, of the same order of magnitude as our experimental result.

\section{Force jump and T1 events.}
\label{force_jump}

The elastic loading is interrupted by sudden force decreases. For young foams, the force jumps distribution is well fitted by an 
exponential law (see Fig. \ref{saut_dis_jeune}). This result questions the universality of the power-law behavior of the stress drop found in 2D Couette geometry. A power law with exponent -0.8, followed by an exponential cut off, was found experimentally by Dennin {\it et al} with a sheared bubble raft \cite{lauridsen02,pratt03}, in good agreement with numerical results obtained by Durian {\it et al} using the bubble model (exponent -0.7)\cite{durian95,durian97,tewari99}. Okuzono {\it et al}, using a vertex model, also found a power law, but with a exponent  -1.5 \cite{okuzono95}.
In our experiment, the force jump distribution is qualitatively modified  when the coarsening process becomes non-negligible and a power law with an exponent close to -1.5 is then observed (see Fig. \ref{saut_dis_jeune}).
The detailed study of the transition between both behaviors will be the aim of a future work. 
This paper only focuses on the non-coarsening foams and the oldest foams, that however have the same shear modulus and plastic threshold than the youngest, are disregarded in the following. 

If the mean force jump $<\delta F>$ is simply computed from the list of recorded  force jumps, its value depends on the arbitrary parameter $\epsilon$, which is the cut off for the small jumps. We thus chose  to extract $<\delta F>$  from the exponential fit, extrapolated to zero. We checked that both methods lead to results variations smaller than the error bar.
The averaged jump amplitude as a function of the bubble size is 
 well fitted by the linear relation  (see Fig. \ref{saut_moy}) :
\begin{equation}
<\delta F> = 1.0 \gamma d  \, \pm 10\% . 
\end{equation}
This scaling law is consistent with the simple approach which identifies the plastic event 
responsible for the force jump with a single bubble detaching from the bead.
Neglecting the contribution of the pressure (which may in fact play a significant role, at least in 2D \cite{dollet05b}), the force exerted on the bead roughly scales as $\cal L^+ - \cal L^- $, with $\cal L^+$ and $\cal L^-$ the total length of Plateau borders touching the sphere respectively on the downstream side and on the upstream side. As detailed on Fig. \ref{detach_bul}, these 
lengths scale as  ${\cal L}^\pm \sim R \sqrt{n^\pm}$, with  $n^\pm$  the numbers of bubbles on each side.  If one bubble detaches from the downstream side, the variation of $\cal L^+$ scales as
$\delta {\cal L^+} \sim - R/\sqrt{n^+} \sim d$ which leads to $\delta F \sim \gamma d$.
The foam behavior can thus be understood, at the first order, with very localized events, which is confirmed by the spatial T1 distribution discussed in the section \ref{spatial_dis}.

 The mean 
force jump duration is $<t_{max,i}-t_{min,i+1}>= 1.5 s \pm 0.5s$. A closer analysis of the force signal indicates that the relevant criterion for the jump beginning
 is the slope modification 
and not the maximum of the force. This is quite difficult to detect automatically and thus an arbitrary duration has been assigned to the force jump $i$, which begins at $t_{max,i}-2 (s)$ and finishes at $t_{min,i+1}+0.5 (s)$. 
With this convention a satisfying pairing has been obtained between the images and the  force measurements, for all the experiments made with bubbles larger than 3mm in diameter (for smaller bubbles no information is obtained from the images).  In average, $85\%$ of the spots occur during the $15\%$ of the time corresponding to force jumps. The 15$\%$ remaining spots probably correspond to small force jumps, smaller than $5 \mu$N, 
of the order of the noise and not taken into account.  
We obtain in average 2.5 plastic events  per force jump. For each force jump, the volumes of these events, determined as explain in section \ref{exp_meas}, 
are summed to get the total volume of foam reorganized during the jump. The Fig.\ref{volT1} shows the correlation 
between the force amplitude and the volume of reorganization. 

Most of the plastic events only appear on one image, which makes the estimation of the event duration difficult.
Nevertheless, the number of successive images on which a given plastic event appears is exponentially distributed, and the average event duration  deduced 
from this distribution is $\tau = 0.2 s$. This time is much smaller than the time scale obtained from the force jump. The Plateau borders appear on the image with a typical size of 5 pixels and it moves from its own 
thickness between 2 images only if the velocity is of the order of 2 mm/s. This large velocity is probably only reached during a small fraction of the whole T1 transformation, thus explaining the difference between the two time scales.

\section{Plastic events spatial distribution}
\label{spatial_dis}

The correlation between the localization of the plastic events and the force jumps amplitudes would be very interesting, and will be feasible with a better jump statistics and an improved determination of the spot position in the 3D space. 
In this paper, we simply discuss the spatial distribution of the plasticity, averaged over the whole measurement time.
This does not require to identify each individual event and thus allows to suppress the thresholding from the image processing and to take advantage of the whole information contained in the grey levels.
After subtracting two successive 
images, we take the absolute value and we  put to zero all the pixels with a grey level smaller than the noise level. The other grey level values, containing the information on the displacements, possibly smaller than one pixel, remain unchanged.  Then all the images are summed. 
After a grey level binning process keeping only 5 different values, only used for a better legibility, we obtain typically the image shown on Fig.\ref{iso}. 

The isodensity lines are organized around the two high stress regions, in front of the bead and behind.
 The experimental data correspond to a projection in the (x,y) plane, but, as the 
plasticity decreases quickly with the distance from the bead center, it has qualitatively the same angular behavior as the distribution in the median plane defined as $z=0$ (the frame orientation is shown on Fig. \ref{iso}). It can thus be directly compared to the maximal shear stress angular distribution in the  $z=0$ plane for a purely elastic medium, as plotted on Fig \ref{iso}.
Both distributions show similarities, but the plasticity lobe upstream ($x>0$) extends further laterally than the other one (independently of the presence of the rod), whereas the elastic shear stress distribution is symmetrical (see Fig.\ref{iso}). 

The radial distribution has been investigated along the line $y = 0$ which has the largest signal/noise ratio. Local fluctuations have been reduced by averaging
the grey level along a 25 pixels segment  oriented in the  ${\bf u}_y$ direction.
The grey level decreases phenomenologically as $k_1 \, \mbox{exp}(-x^2/k_2^2)$, as depicted on Fig.\ref{coupe}. As the projection only induces
polynomial transformations of the signal, the exponential decreases also applies in the 3D space, along the line (y=z=0). 
In a purely elastic medium, the stress amplitude decreases as $1/x^2$ along the same line (see eq.\ref{stress_tensor}).

The typical distance over which the plasticity disappears is of the order of $2 d \pm 0.3$ and seems to be slightly larger 
in presence of the rod. The plastic events are thus almost all in the first and second layers of bubbles and the foam structure is not strongly modified further. Leonforte {\it et al} performed numerical simulations in a 3D Lennard-Jones fluid in a similar geometry, in the linear regime \cite{leonforte05}. They demonstrate that the average stress field is well described by the classical 
elasticity in this case, but that the stress fluctuations decrease exponentially when the distance to the bead increases. This may be related 
to the exponential decrease we obtained experimentally, but the characteristic distances obtained numerically are much larger, of the order of 20 interatomic distances.
Exponential decrease of the plastic events density has already been observed in 2D foams in Couette geometry, with a typical distance of the order of few bubble layers
\cite{debregeas01}.

\section{Conclusion}

This article reports on a Stokes flow experiment performed in a 3D dry foam, in a quasi-static regime. 
We provide an original measure of the foam shear modulus and of the plastic threshold, and show that a simple model 
of elasto-plastic material allows to predict the scaling laws obtained for the mean force exerted on the bead, as well  as for the average value of the force fluctuations. The structure reorganizations
are recorded and their characteristic volumes are correlated to the amplitude of the force jumps they induce.
Finally we show that the plastic events density decreases exponentially with the distance to the bead, 
as it is typically observed in case of shear banding in foam \cite{debregeas01} or in granular matter \cite{pouliquen96}.

In conclusion, this experiment is an interesting alternative to the more classical simple shear geometry to probe the 
mechanical response of the foam. As the force is localized, only a small amount of bubbles are involved and the stress and strain fluctuations are more pronounced. Additionally, there is no need to suppress the sliding along the wall : the bead surface is smooth and thus induces no external random noise. 
The preliminary results concerning the influence the coarsening on the distribution of the force fluctuations and on the plasticity localization will be studied in greater details in a future work.

\vspace*{0.5cm}

{\bf Acknowledgments}\\
The authors are grateful to the French Spatial Agency (CNES) and to the CNRS for financial support. We thank Mr Hautemayou and Mr Laurent for technical help. IC thanks the LPMDI for its hospitality. 

\section*{Appendix A}
For an elastic medium of Lame parameters $\lambda$ and $G$, 
the constitutive relation between the stress tensor $\sigma_{ij}$ and the deformation tensor
 $\epsilon_{ij}= 1/2 (\partial_i X_j + \partial_j X_i)$, with $\partial_i$ the first derivative with respect to the $i^{th}$ spatial coordinate,
is 
\begin{equation}
\sigma_{ij}= \lambda Tr(\epsilon) + 2 G \epsilon_{ij}
\end{equation}
The equilibrium equation, without external force is given by
\begin{equation}
 0= G \sum_j \partial_{jj} X_i + (\lambda+ G) \partial_i (\mbox{div} {\bf X})
\end{equation}
In the incompressible limit, $\lambda \rightarrow \infty$, it becomes
\begin{equation}
  0= G \sum_j \partial_{jj} X_i - \partial_i P  \hspace{0.5cm}, \hspace{0.5cm} 0 = \mbox{div} {\bf X}
\end{equation}
with $ P = -1/3 Tr(\sigma)= -(\lambda+2 G/3)Tr(\epsilon)$

These equations are analogous to the hydrodynamical equations in the viscous regime.
The displacement field around a sphere of radius R displaced from its equilibrium position on the distance $\delta_x$ in an 
infinite 3D incompressible elastic medium is given, in spherical coordinates $(r, \theta, \phi)$, by \cite{guyon}
\begin{equation}
 X_r = \delta_x \cos \theta {R \over r}\hspace{0.5cm}, \hspace{0.5cm} X_{\theta}= - \delta_x \sin \theta {R \over 2 r}
\end{equation}
The direction ${\bf u}_x$ of the bead motion is the polar axis of the spherical coordinates and the bead center is the 
frame origin.
A vanishing tangential force is assumed on the sphere.
The resulting stress tensor is 
\begin{equation}
 \sigma= - P I + G \delta_x {R  \cos \theta \over r^2} \left[\begin{array}{rrr}-2 & 0 & 0 \\ 0& 1&0 \\ 0&0&1 \end{array} \right]
\label{stress_tensor}
\end{equation}
The maximal shear stress value, relatively to the orientation of the normal vector, is 
\begin{equation}
\sigma_{sh}=1.5 \, G \delta_x { R  \cos \theta \over  r^2 } 
\label{shear_stress_max}
\end{equation}
 and the resulting force on the sphere is 
 \begin{equation}
F= 4 \pi G R \delta_x
\end{equation}

\newpage

\begin{table}[h]
\caption{\label{table_mod} Previous experimental and theoretical results for foams or emulsions shear modulus in the dry limit.}
\begin{ruledtabular}
\begin{tabular}{lccc}
 &Theo./Exp./Num  & Set-up 
 & $G / (\gamma/r_V)$ \\
\hline
Derjaguin\cite{derjaguin33}& T   & & 0.89  \\
Derjaguin\cite{derjaguin34}& E, foam & rheometer osc. & 0.8 \\
Stamenovi\'c\cite{stamenovic84}& E, foam   & rheometer stat.& 0.73  \\
Princen\cite{princen86}& E, emuls.  & rh. stat. & 0.51  \\
Budiansky\cite{budiansky91}  & T   & & 0.59  \\
Stamenovi\'c\cite{stamenovic91}& T   & & 0.55  \\
Sun\cite{sun94}& E, foam  & Wave prop. & 1.69 \\
Mason\cite{mason95}& E, emuls.  &  rh. osc. & 0.6 \\
Coughlin\cite{coughlin96}& E, foam  & Indentation & 0.4  \\
Kraynik\cite{reinelt96,kraynik96}& N, Kelvin C. & Surf. Evol. & 0.35-0.60  \\
Saint-Jalmes\cite{saintjalmes99}& E, foam &  rh. osc. & 0.51\\
C. and P. & E, foam & Stokes & 0.46 \\
\end{tabular}
\end{ruledtabular}
\end{table}

\newpage
\centerline{Figure captions}
{\bf Fig. 1} Experimental set-up. \\

{\bf Fig. 2} Typical force graph obtained with the foam moving at 5 $\mu$m/s. The box is moving downwards, so the force exerted by the foam on the bead
is oriented  downwards too and the balance signal is positive. The transient finishes at $t=1000 s$. The open circles on the magnified graph represent the maxima and minima recorded and the dots are the measurement points. \\

\begin{figure}[h]
\includegraphics[width=7cm]{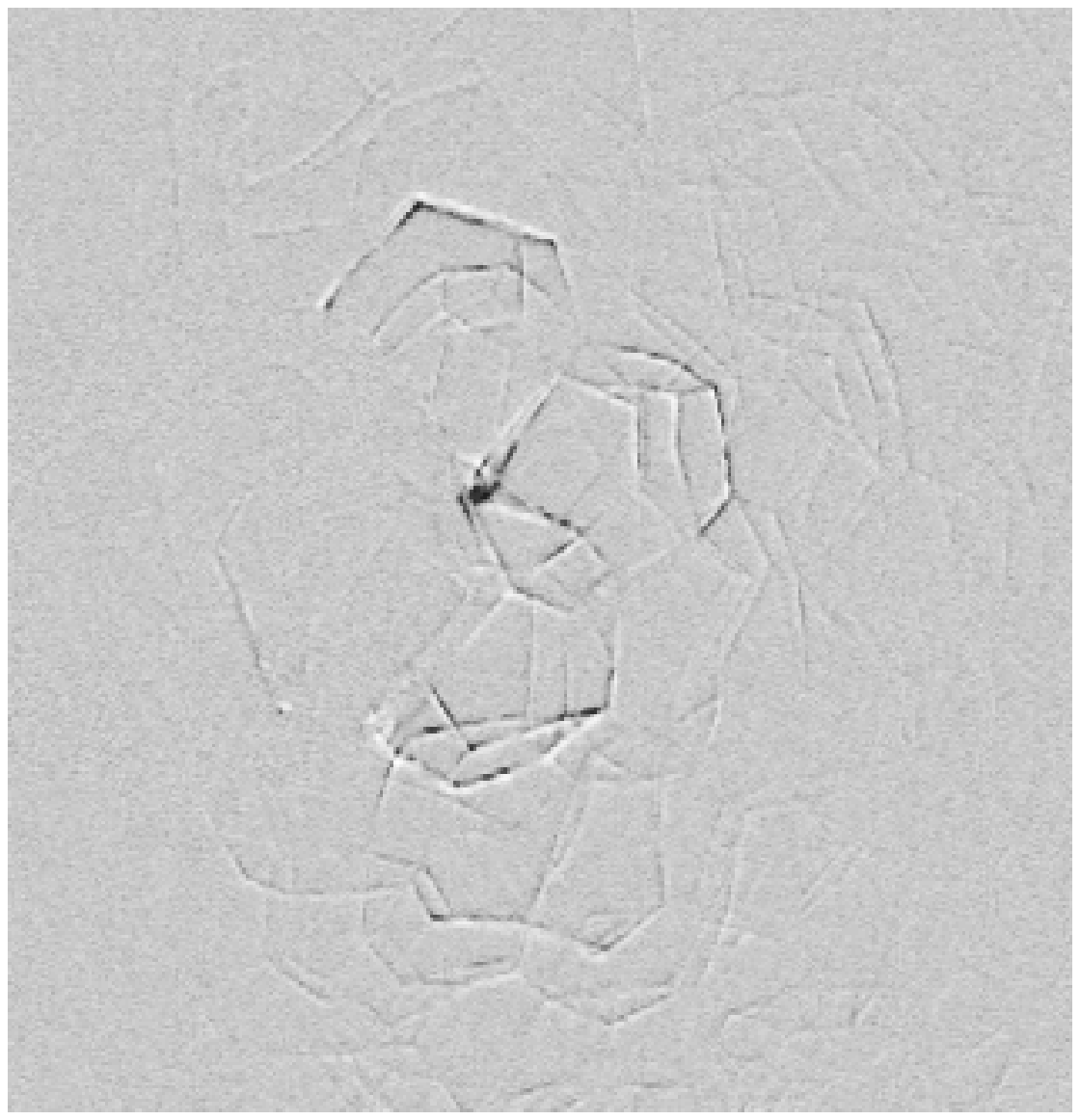}		
\includegraphics[width=7cm]{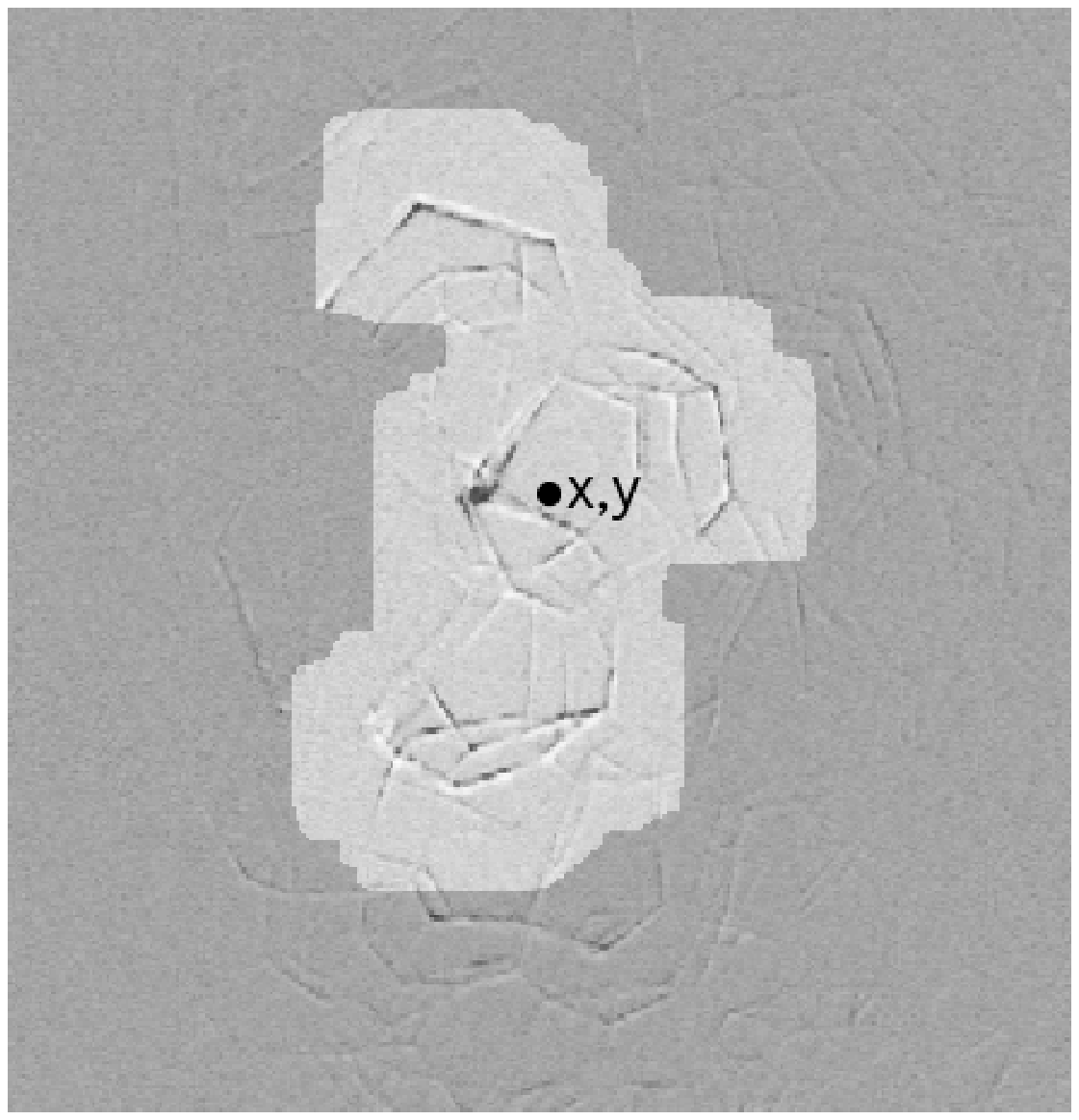}
\caption{Cantat, Physics of Fluids.} 
\label{diff_image}
\end{figure}

\begin{figure}[h]
\includegraphics[angle=-90,width=11cm]{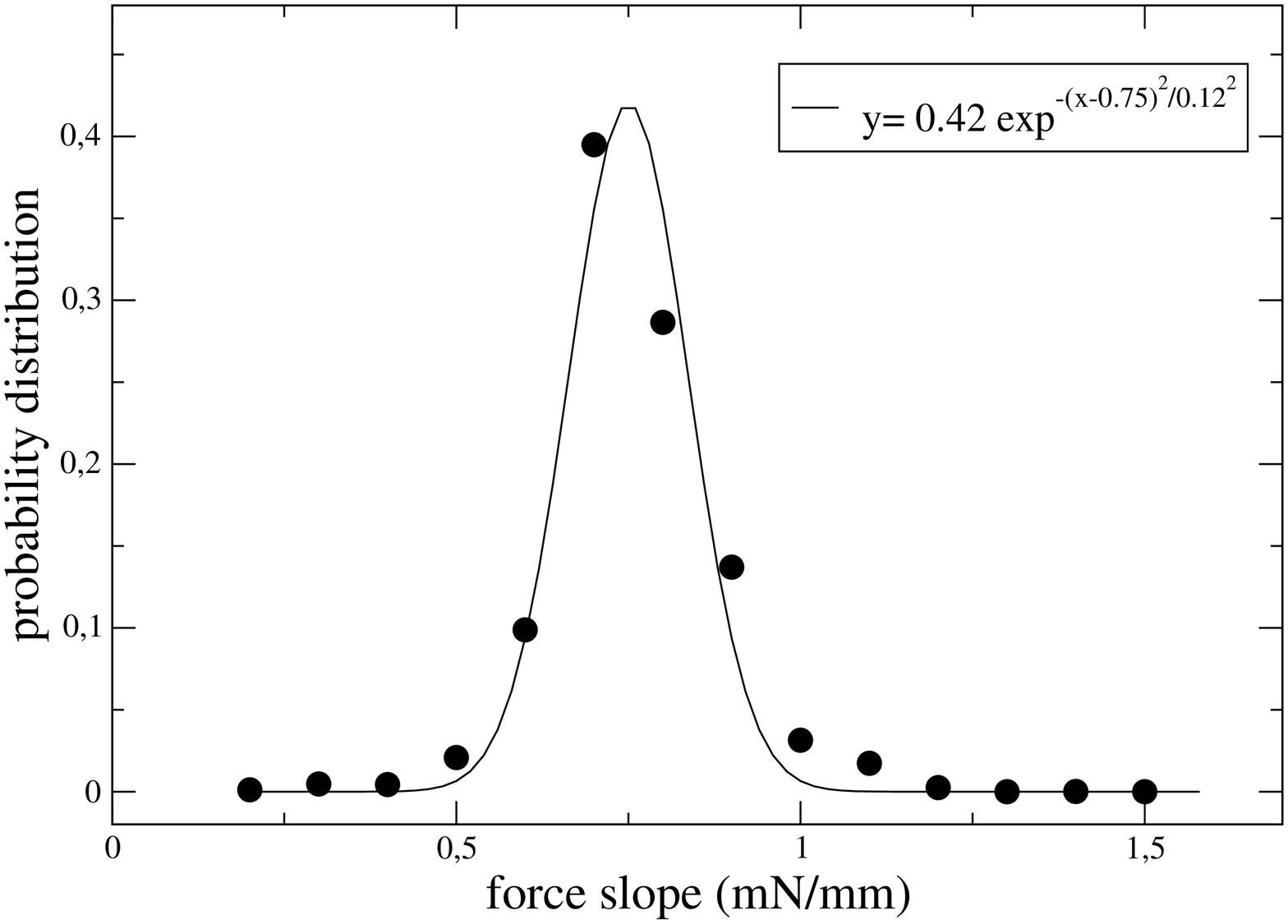} 
\caption{Cantat, Physics of Fluids.}
\label{pente_dis}
\end{figure}

\begin{figure}[h]
\includegraphics[angle=-90,width=11cm]{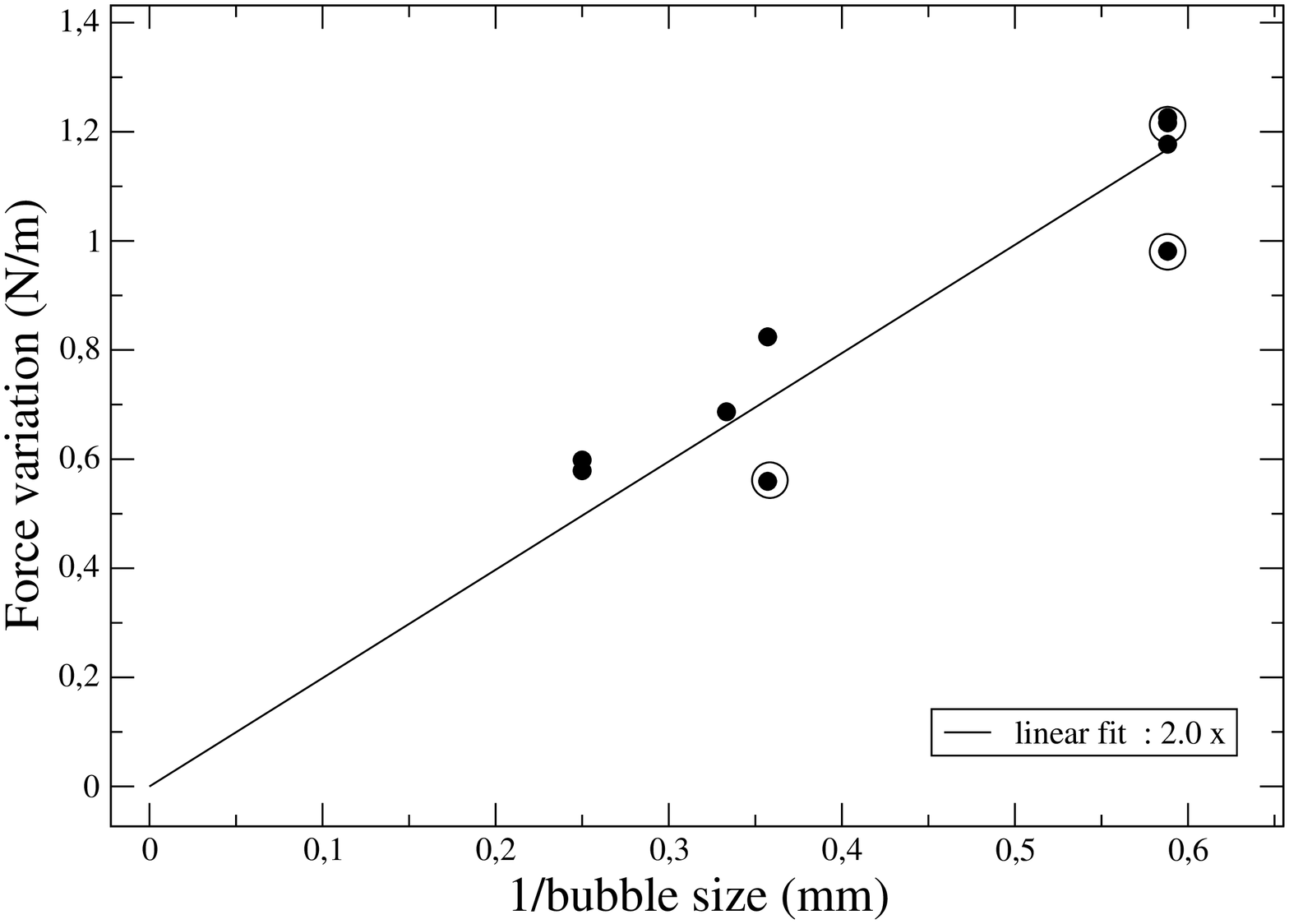}
\caption{Cantat, Physics of Fluids.}
\label{pente}
\end{figure}

\begin{figure}[h]
\includegraphics[angle=-90,width=11cm]{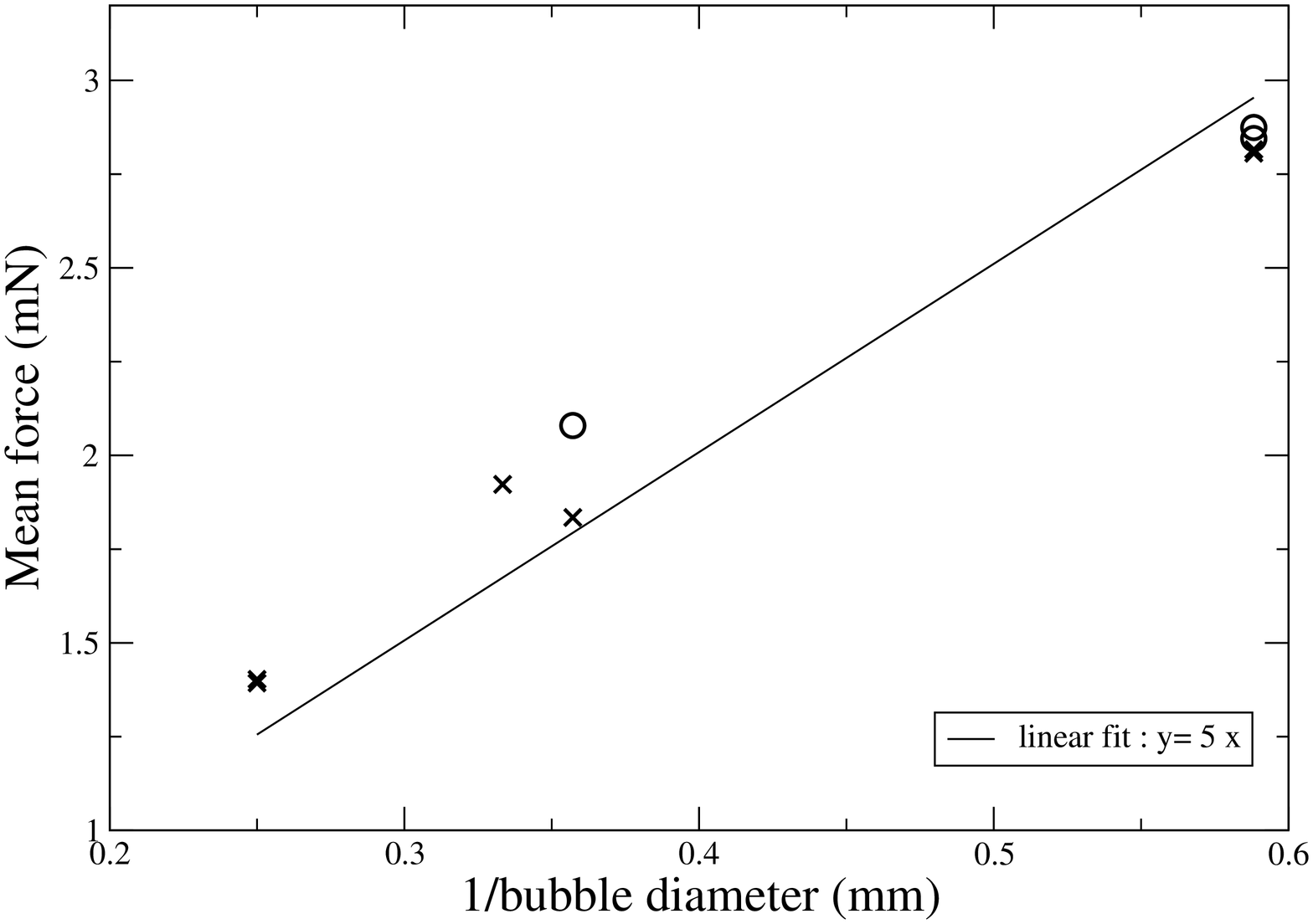}
\caption{Cantat, Physics of Fluids.}
\label{force_moyenne}
\end{figure}

\begin{figure}[h]
\includegraphics[angle=-90,width=11cm]{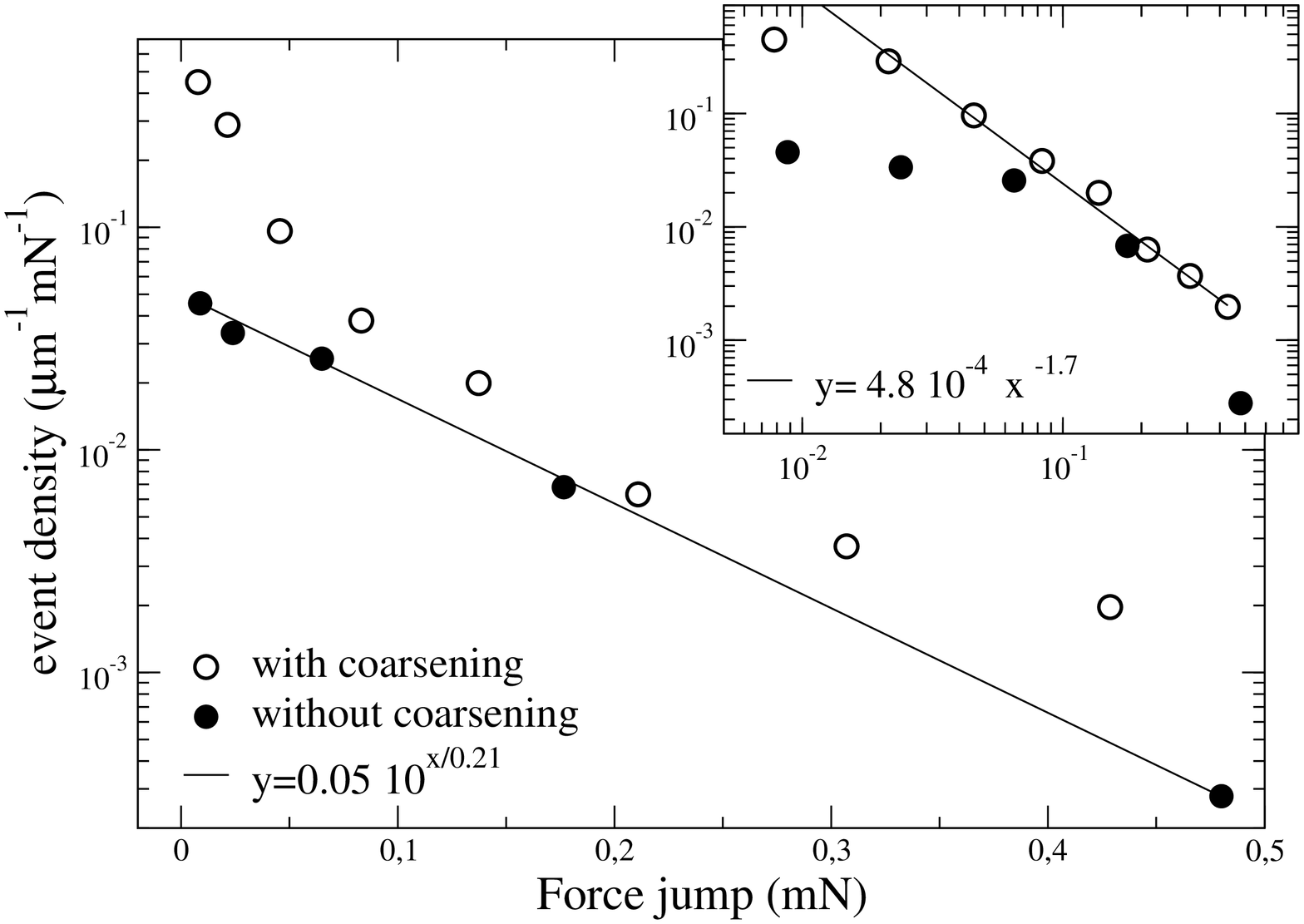}
\caption{Cantat, Physics of Fluids.}
\label{saut_dis_jeune}
\end{figure}

\begin{figure}[h]
\includegraphics[angle=0,width=11cm]{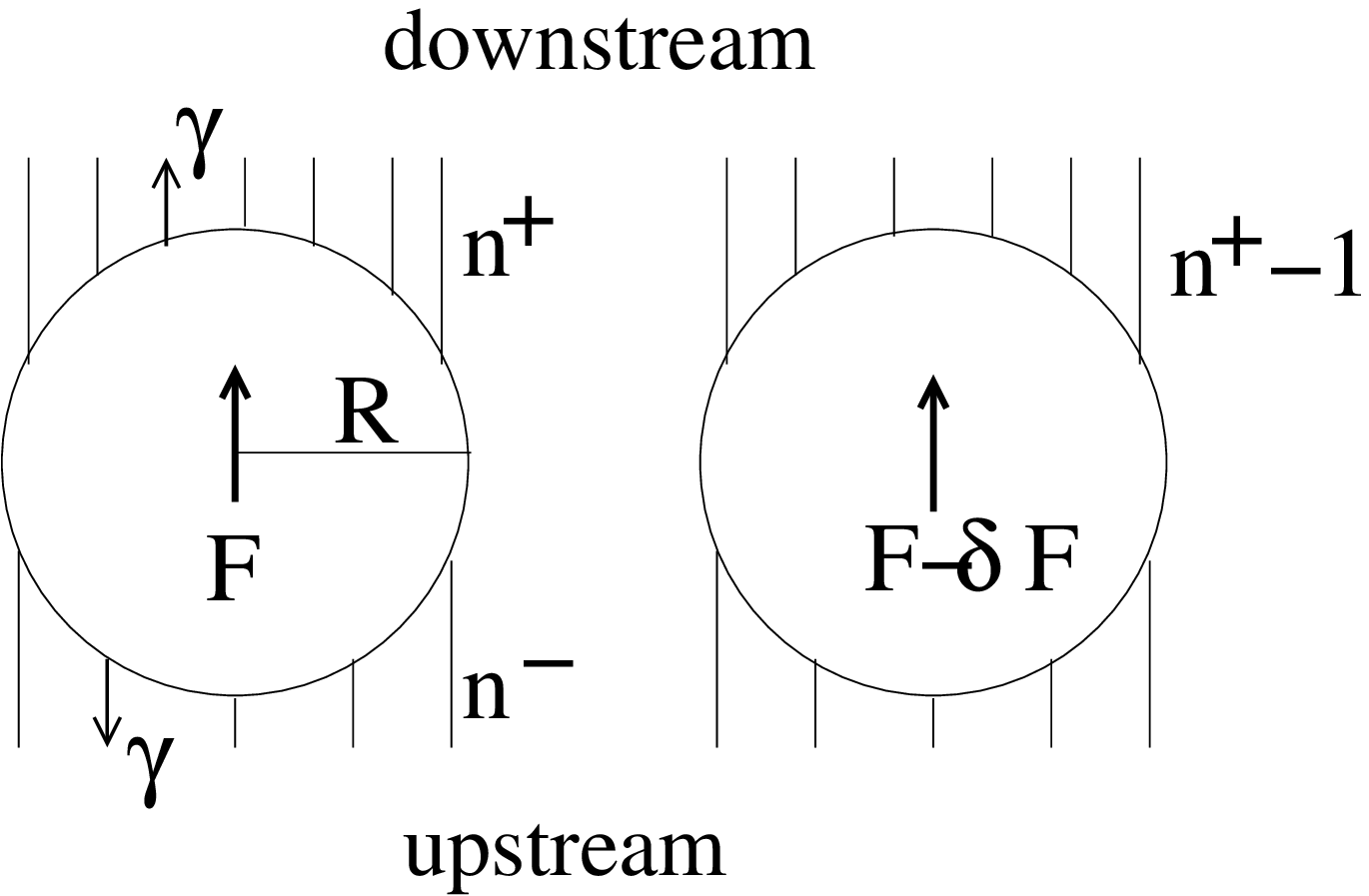}
\caption{Cantat, Physics of Fluids.}
\label{detach_bul}
\end{figure}

\begin{figure}[h]
\includegraphics[angle=-90,width=11cm]{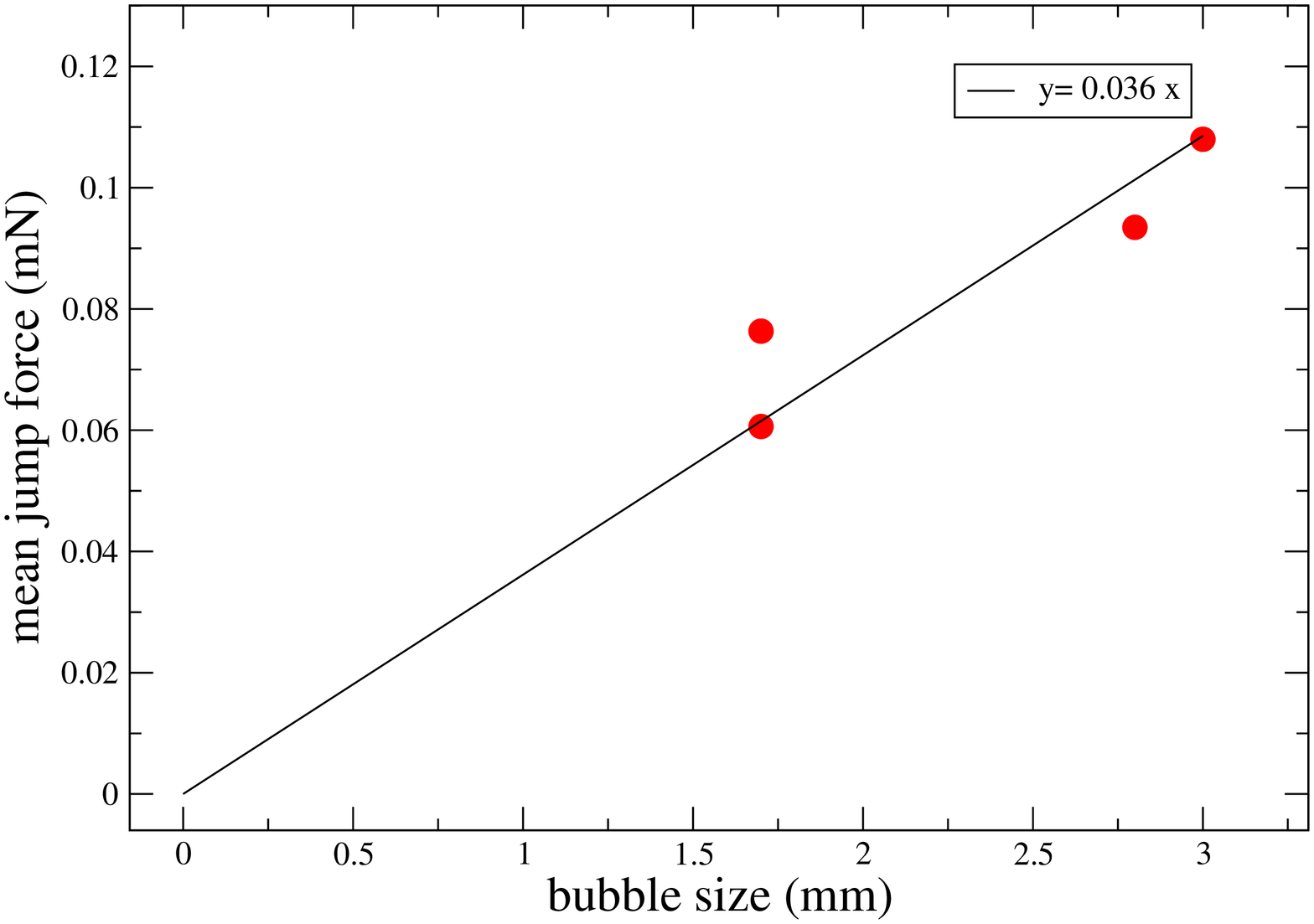}
\caption{Cantat, Physics of Fluids.}
\label{saut_moy}
\end{figure}

\begin{figure}[h]
\includegraphics[angle=-90,width=11cm]{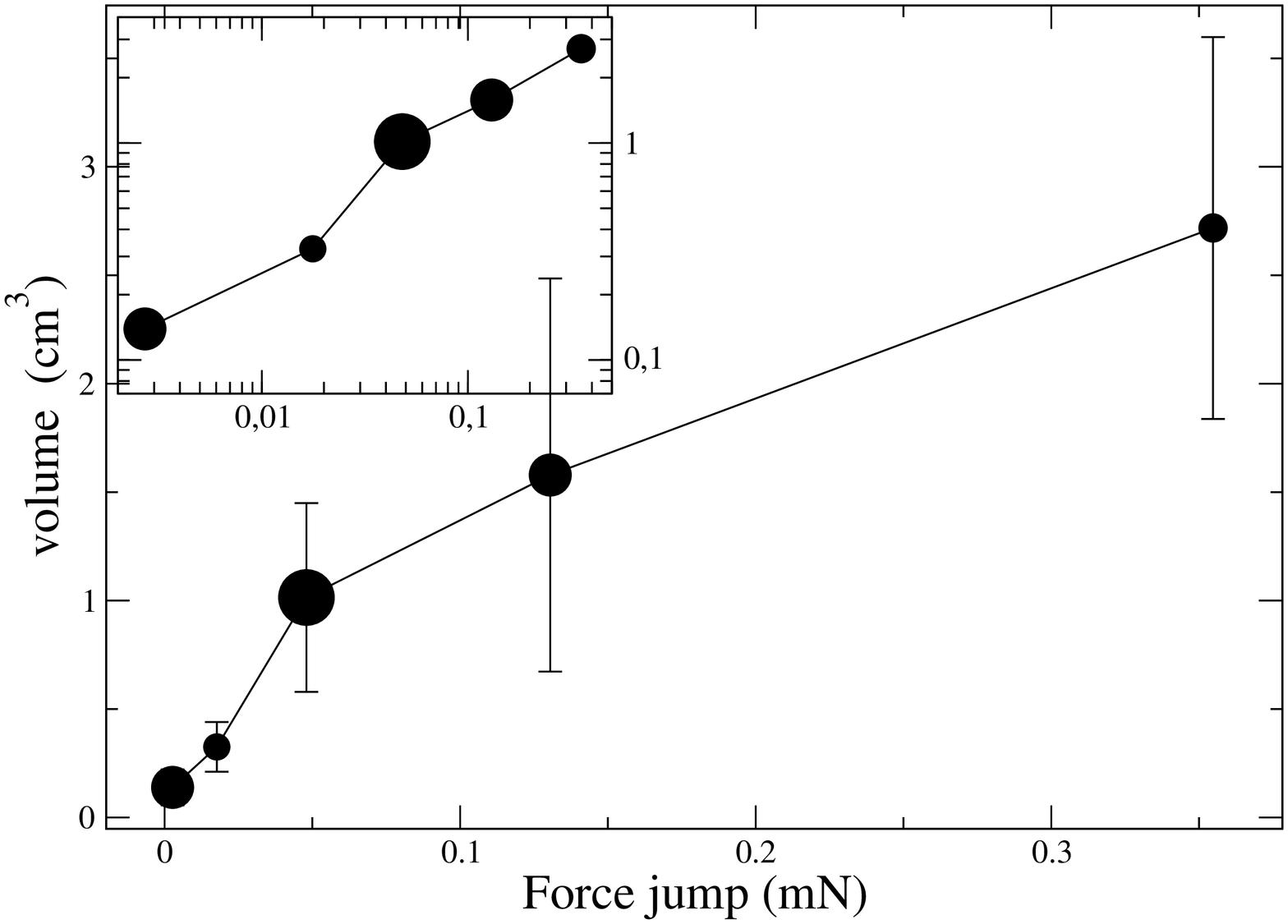}
\caption{Cantat, Physics of Fluids.}
\label{volT1}
\end{figure}

{\bf Fig. 3} Raw image obtained from the camera. The small rectangle is shown on Fig.\ref{diff_image} at two stages of the image processing.\\

{\bf Fig. 4} A T1 event as it appears on the difference between two successive images (left), and after
image processing (right). The value retained to measure the plastic event volume is the area of the spot to the power 3/2.\\

{\bf Fig. 5} Probability distribution of the force variation with the bead displacement in a 3mm bubbles foam, during the elastic loading. It is obtained from the slopes of all the smooth force increases of a typical flow, weighted by the force increase duration. The events number is 162.  The full line is the best Gaussian fit.\\

{\bf Fig. 6} Mean value of the force slope as a function of the bubble size $d$. The simple dots and the dots with a circle correspond respectively to an upwards and to a downwards motion of the bead. No significant difference is observed between both 
directions of motion, so the asymmetry due to the rod is less than our error bar (of the order of $15\%$). The full line is the best fit of the type y=ax.\\

{\bf Fig. 7}  Force exerted by the foam on the bead, averaged over the steady state. (o) : downward motion of the box, ($\times$) : upward motion.  The solid line is a linear fit.\\

{\bf Fig. 8} Force jumps distribution for a 45 min old foam with 2.8mm bubbles (no coarsening, 68 events) and for a 2h old foam without $C_6F_{14}$ (coarsening, 251 events) in semi-log and log-log (in insert) representation.
The event density $\rho_{\mbox{jump}}(A)$ is related to the number of force jumps $n_{\mbox{jump}}$ which have amplitude comprised between $A-\Delta A/2$ and $A+\Delta A/2$, observed during a 1 micron bead displacement, by $\rho_{\mbox{jump}}(A) \Delta A = n_{\mbox{jump}}$. Only the force jumps larger than 10 $\mu$N are recorded. We used a non-constant binning to ensure that every point represents approximatively the same events number. 
The horizontal error bar is thus the distance between two points in the x-direction.
  A power law fit seems to be more appropriate for the 
coarsening foam ($\rho_{\mbox{jump}}(A) \sim A^{-1.7 \pm 0.2}\mu$m$^{-1} $mN$^{-1}$  with A in mN), whereas an exponential decay well agrees for the 
non-coarsening foam ($\rho_{\mbox{jump}}(A) \sim 10^{A/0.21\pm 0.05}$ with A in mN). The error bars on the exponents are deduced from the extremal slopes compatible with the horizontal error bars on the data points. The first point is disregarded for the coarsening foam. 
 \\

{\bf Fig. 9} Schematical representation of the force variation induced by a plastic event.  $n^\pm$ are the numbers of bubbles on each sphere side (upstream and downstream) and $d^\pm$ are the typical diameters of their contacts with the sphere. The  relations $n^\pm (d^\pm)^2 \sim R^2$ and $n^\pm d^\pm \sim \cal L^\pm$ leads to the relation ${\cal L}^\pm \sim R \sqrt{n^\pm}$ used in the text.\\

{\bf Fig. 10}  Mean jump force computed from an exponential fit of the force jump distribution, extrapolated to 0, in order 
to take into account the very small jumps, despite the noise. The data are in good agreement with a  linear fit (full line).\\

{\bf Fig. 11} Average volume of foam reorganized during a force jump as a function of the jump amplitude, for a 3mm bubble foam. 
The averages are performed after a force jumps binning and the number of jumps corresponding to each point of the graph is proportional to the circle radius (110 force jumps were recorded in total).
For each force jump amplitude, the standard deviation of the distribution of reorganized volumes is depicted by the error bars.
The insert shows the same data in a log-log plot.\\

{\bf Fig. 12} Spatial repartition, projected in 2D, of the foam quick displacements, averaged during a 1cm displacement of the box, upwards ($-{\bf u}_x$ direction).
The averaging process is discussed in the text. Light and dark grey levels represent respectively the high and low density of plasticity. The full lines are the curves of constant maximal shear stress value in the median plane of a 3D elastic medium, obtained from eq. \ref{shear_stress_max}.\\

\begin{figure}[h]
\includegraphics[angle=0,width=11cm]{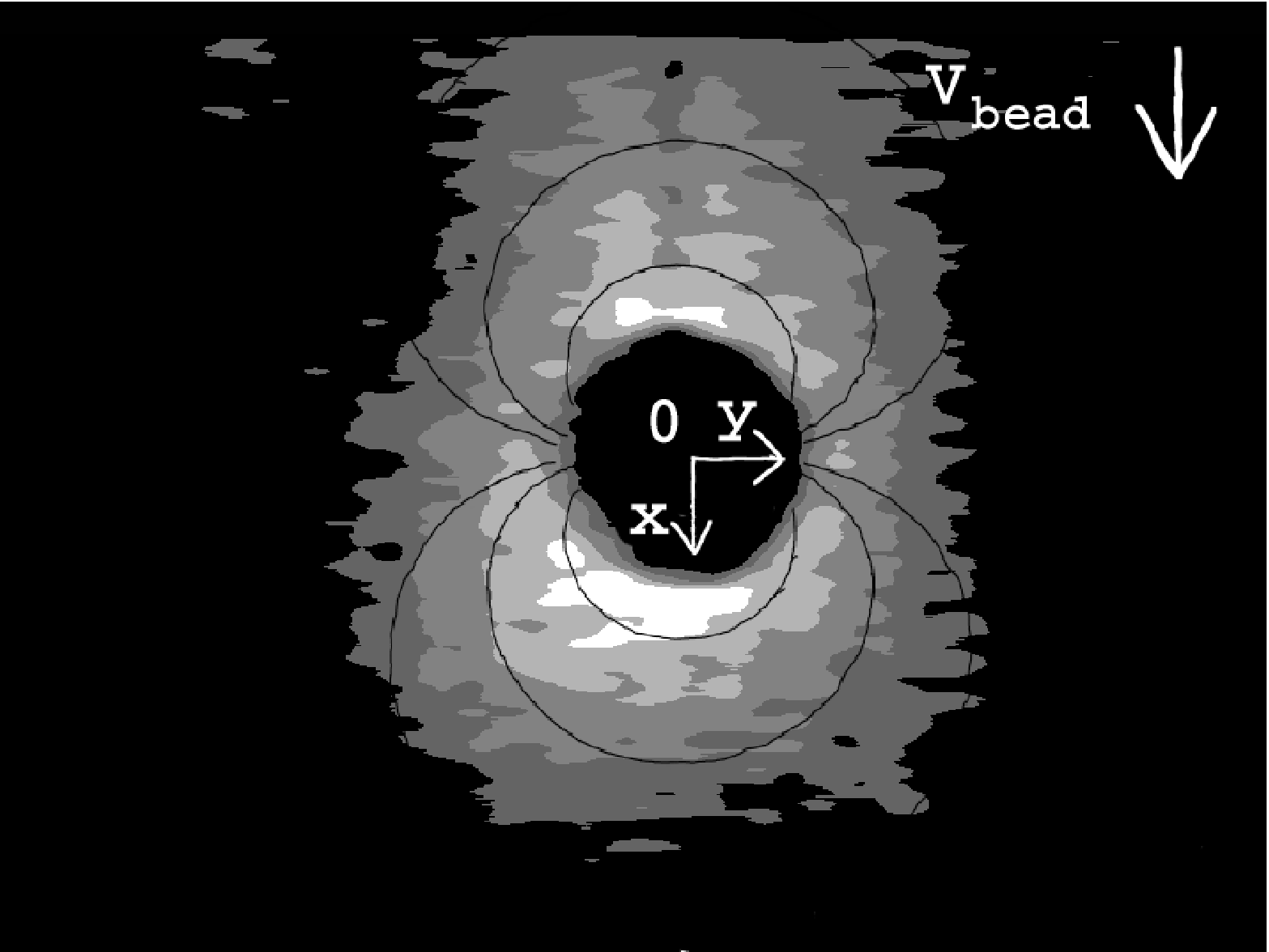}
\caption{Cantat, Physics of Fluids.} 
\label{iso}
\end{figure}

\begin{figure}[h]
\includegraphics[angle=-90,width=11cm]{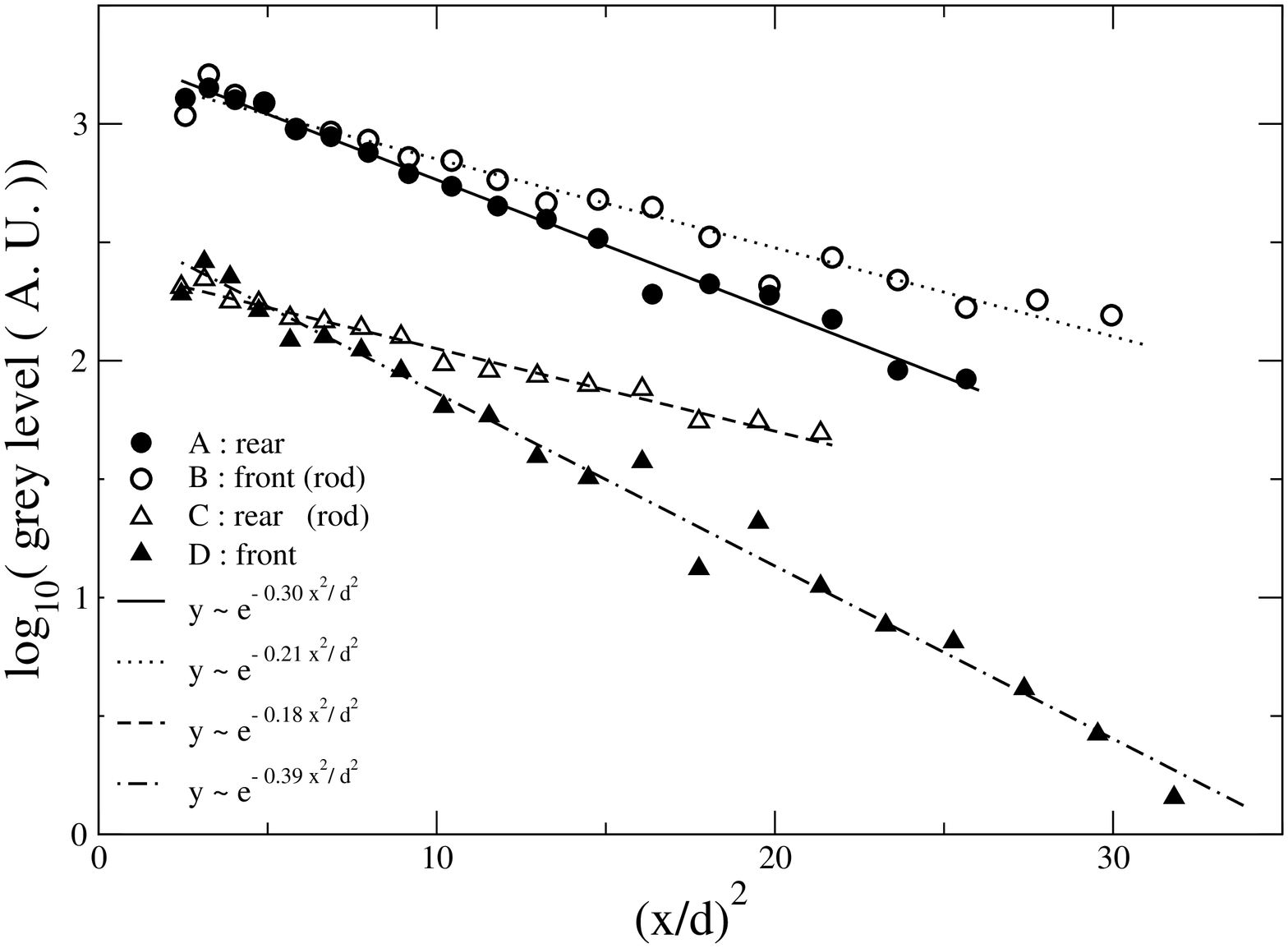}
\caption{Cantat, Physics of Fluids.}  
\label{coupe}
\end{figure}

{\bf Fig. 13}  Radial variation of the plastic events density. 
Series C and D : The grey levels shown on Fig.\ref{iso} are averaged over 25 pixels on the lateral direction, along two vertical lines passing through the bead center, in front of and behind the bead (before the grey level binning). 
The resulting grey levels are plotted as a function of $(x/d)^2$, which underlines the good agreement with the phenomenological adjustment $k_1 \, \mbox{exp}(-x^2/k_2^2)$, with $k_1$ an arbitrary prefactor and  $k_2$ the typical distance 
over which the plastic events density decreases. The origin for $x$ is the center of the sphere.
Series A and B : same kind of data for a foam moving in the other direction, shifted vertically for a better legibility. 

\newpage

\newpage

\newpage

\newpage

\newpage

\newpage

\newpage

\newpage

\newpage

\newpage

\newpage

\newpage

\newpage

\end{document}